\documentclass[a4paper,11pt]{article}
\usepackage[top=3cm,bottom=3.4cm,left=2cm,right=2cm]{geometry}

\usepackage{bm}
\usepackage{amsmath}
\usepackage{amssymb}
\usepackage{cite}
\newcommand{\nn}{\nonumber\\}

\renewcommand{\thepage}{}
\makeatletter
\@addtoreset{equation}{section}
\renewcommand{\theequation}{\thesection.\@arabic\c@equation}
\makeatother
\renewcommand{\thefootnote}{\fnsymbol{footnote}}

\begin{document}
\begin{titlepage}
\title{
\vspace*{-4ex}
\hfill
\begin{minipage}{3.5cm}
\end{minipage}\\
 \bf 
Vertex Operators in Superstring Theory from Integral Forms and Descent Equations
\vspace{1em}
}

\author{
Isao~{\sc Kishimoto},$^{1}$\footnote{\tt ikishimo@rs.socu.ac.jp}
~~~Shigenori {\sc Seki},$^{2}$\footnote{\tt shigenori.seki@setsunan.ac.jp}
~~~Haruka {\sc Shimogaki},$^{3}$\footnote{\tt shimogaki@asuka.phys.nara-wu.ac.jp}
\\
\vspace{-3ex}\\
~and~ 
\\
\vspace{-3ex}\\
Tomohiko~{\sc Takahashi}$^{3}$\footnote{\tt tomo@asuka.phys.nara-wu.ac.jp}
\\
\vspace{0ex}\\
\\
$^{1}${\it
 Center for Liberal Arts and Sciences, Sanyo-Onoda City University,}\\
{\it Daigakudori 1-1-1, Sanyo-Onoda, Yamaguchi 756-0884, Japan}
\vspace{1.5ex}
\\
$^{2}${\it Institute for Fundamental Sciences, Faculty of Science and Engineering,}\\
{\it  Setsunan University,}\\
{\it  17-8, Ikedanaka-machi, Neyagawa, Osaka 572-8508, Japan}
\vspace{1.5ex}
\\
$^{3}${\it Department of Physics, Nara Women's University,}\\
{\it Nara 630-8506, Japan}
\vspace{1ex}
}

\date{}
\maketitle

\begin{abstract}
\normalsize
We develop a geometric formulation of vertex operators in superstring
theory based on integral forms on super Riemann surfaces. Starting from
the integrated NS-NS vertex operator, we derive descent equations that
relate operators with different ghost and picture numbers.
A key result is a correspondence between
supergeometric objects and ghost superfields, in which the one-form
$dz-\theta d\theta$ and the even differential $d\theta$ are identified with
the ghost superfield and its superderivative. This provides a
geometric realization of the superghost structure.
We further extend the construction by incorporating inverse picture-changing operators, which
generate new descent sequences across different picture sectors.
We also introduce
a superfield construction of higher-ghost-number
operators, for which additional terms are required compared to the bosonic
case.  
All operators are
organized into a universal descent structure and are well-defined in
BRST cohomology. 
\end{abstract}

\end{titlepage}

\renewcommand{\thepage}{\arabic{page}}
\renewcommand{\thefootnote}{\arabic{footnote}}
\setcounter{page}{1}
\setcounter{footnote}{0}
\section{Introduction}

In the BRST formalism, physical states of superstring theory are
identified as elements of the BRST cohomology: states that are
annihilated by the BRST operator but are not BRST exact. In the
Neveu--Schwarz Neveu--Schwarz (NS-NS) sector, vertex operators
representing such states can be realized in different pictures,
reflecting the structure of the superghost system,
as originally formulated in~\cite{Friedan:1985ge}
and discussed in Secs.~12.3 and 12.5 of~\cite{Polchinski:1998rr}.
Although the physical
content of the theory is independent of the picture number, the explicit
form of vertex operators and their insertion rules do depend on it.
Picture-changing operators, together with
their inverses, relate different picture representations while preserving
the BRST cohomology~\cite{Friedan:1985ge}.

From a geometric perspective, vertex operators can be interpreted as
operator-valued differential forms on the worldsheet,
or more generally in superspace. This viewpoint naturally leads to a
hierarchical structure governed by descent equations involving the BRST
differential and the exterior derivative. In bosonic string theory, such
descent equations provide a systematic relation between integrated and
unintegrated vertex operators and allow the construction of operators
with various ghost numbers~\cite{Kishimoto:2024yuw}. These developments
offer a powerful framework for organizing vertex operators by ghost
number and form degree.

In superstring theory, however, a comparable framework that
simultaneously incorporates picture number, superghosts, and superfields
in a geometric manner has been lacking.
In this paper, we develop such a framework by extending
the descent formalism to superspace using superfields,
and expressing it in the language of supergeometry and integral forms
~\cite{Voronov:1986a,Voronov:1986b,Voronov:1987,Voronov:1988,Belopolsky:1996cy,Belopolsky:1997bg,
Belopolsky:1997jz,Witten:2012bg}.
The resulting formulation clarifies the underlying algebraic and geometric
structure of the NS-NS sector and provides a unified geometric description of
vertex operators across different ghost and picture sectors.
The resulting framework is expected to have applications in
broader contexts, including higher-genus amplitudes,
loop computations, and extensions to the Ramond sector.

In this context, it is useful to recall that,
in the bosonic string, the geometric interpretation is closely tied to a
natural correspondence between differential forms and ghost fields,
such as $dz \leftrightarrow c$.
It is then natural to expect that, in the superstring case,
this correspondence extends naively to
$dz \leftrightarrow c$ and $d\theta \leftrightarrow \gamma$,
reflecting the structure of the superghost system.

However, we will show that the correct correspondence is more subtle and
cannot be captured at the level of component fields,
and is naturally formulated at the level of superfields.
In particular, the geometrically natural one-form $dz-\theta d\theta$,
and the even differential $d\theta$ on $\varPi TM$ are related to the ghost superfield
$C(z,\theta)$
and its superderivative as
\begin{align*}
dz - \theta d\theta \,\sim\, C,
\qquad
d\theta \,\sim\, -\frac{1}{2}DC.
\end{align*}
This relation will be derived by applying the geometrically defined
picture-changing operators to the descent equations in superspace.

This correspondence should not be understood at the level of component fields.
Rather, it is intrinsically a superfield-level relation and cannot be
consistently formulated as a componentwise identification,
as is also suggested by the geometric framework of supermanifolds
reviewed in~\cite{Witten:2012bg}.
This refined identification provides the geometric origin of
superghost insertions and plays a central role in our geometric formulation
and in the structure of the descent equations in superspace.

This paper is organized as follows.  In section~\ref{sec:supergeometry},
we present the geometric framework of supermanifolds and integral forms
that underlies our construction.  Following \cite{Witten:2012bg}, we
describe differential forms as functions on the parity-reversed tangent
bundle and then introduce integral forms.  We also explain, following
\cite{Belopolsky:1997bg,Belopolsky:1996cy,Belopolsky:1997jz}, how
picture-changing operators arise in the supergeometric framework.  In
section~\ref{sec:NSNSvertex}, we construct vertex operators in the NS-NS
sector within the geometric framework.  Starting from the top integral
form representing the integrated vertex operator, we analyze its BRST
transformation and derive the associated descent equations.  We then
apply the picture-changing operators defined geometrically in
section~\ref{sec:supergeometry}.  In particular, we recover the standard
unintegrated vertex operator in a purely geometric manner, and show that
this procedure naturally leads to a correspondence between
supergeometric objects and ghost superfields.  In
section~\ref{sec:IPCO}, we extend the construction by incorporating
inverse picture-changing operators in the BRST formalism.  In
section~\ref{sec:gh3}, we further extend the construction to vertex
operators with higher ghost number.  We derive the corresponding descent
equations and show that, in contrast to the bosonic case, additional
terms are required for consistency.  We also analyze their
superconformal transformation properties and demonstrate that the
resulting operators are well-defined in BRST cohomology.  Section~6
offers concluding remarks and outlines possible extensions of this work.

\section{Geometric framework of integral forms on
super Riemann surfaces\label{sec:supergeometry}}

To formulate the descent equations for NS-NS vertex operators,
we review the geometric framework of supermanifolds and
integral forms.
The geometric formulation of integration on supermanifolds
was originally developed in a series of works~\cite{Voronov:1986a,
Voronov:1986b,Voronov:1987,Voronov:1988},
and later reviewed in a form suitable for superstring theory
in~\cite{Witten:2012bg}.
In this formulation,
differential forms are described as functions
on the parity-reversed tangent bundle and naturally extended
to integral forms.

In this framework, the notion of picture
is associated with the number of delta-function insertions of even
differentials, and thus differs from the conventional picture number
in the BRST formulation of superstring theory.
The relation between these two notions has been extensively discussed
in~\cite{Belopolsky:1997jz}, where picture-changing
operators are constructed both in supergeometry and in the BRST
complex, and a correspondence between the two is established.
In the present work, however, we treat these two notions as
distinct.

\subsection{Differential geometry on the parity-reversed tangent bundle}

Let $M$ be a supermanifold of dimension $(m|n)$ with local coordinates
\begin{align}
x^A=(x^a,\theta^\alpha),
\end{align}
where $x^a$ ($a=1,\dots,m$) are bosonic coordinates and
$\theta^\alpha$ ($\alpha=1,\dots,n$) are fermionic coordinates.
The collective index $A$ runs over both types of coordinates.
The Grassmann parity is denoted by $|\cdot|$, so that
\begin{align}
|x^a|=0, \qquad |\theta^\alpha|=1 .
\end{align}

The tangent bundle $TM$ carries coordinates $(x^A,dx^A)$,
where the parity of $dx^A$ coincides with that of $x^A$.

A central role is played by the parity-reversed tangent bundle,
denoted by $\varPi TM$.  This is obtained from the tangent bundle $TM$
by reversing the Grassmann parity of the fiber coordinates.
Thus the coordinates on $\varPi TM$ are $(x^A,dx^A)$ with
\begin{align}
|dx^A| = 1-|x^A|.
\end{align}

Functions on $\varPi TM$ are naturally identified with differential
forms on $M$, so that the algebra of differential forms
can be regarded as the algebra of functions
$C^\infty(\varPi TM)$, where $dx^A$ correspond to
basis one-forms. Note that
$dx^a$ are odd, while $d\theta^\alpha$ are even.

In this supergeometric description, the exterior derivative,
contraction, and Lie derivative are realized as
differential operators acting on functions on $\varPi TM$.

As a simple example, 
consider $M=\mathbb{C}^{1|1}$ with coordinates $(z,\theta)$. 
On $\varPi TM$, the coordinate $dz$ is odd while $d\theta$ is even. 
A generic function on $\varPi TM$ then takes the form
\begin{align}
f_0(z,\theta)+ f_1(z,\theta) dz
+ \sum_{k\ge1}
\left(
g_0^{(k)}(z,\theta)+
g_1^{(k)}(z,\theta) dz
\right)(d\theta)^k .
\end{align}
Since $d\theta$ is even, this is not nilpotent,
and therefore
ordinary differential forms do not terminate at a finite degree.
This motivates the introduction of integral forms.

On $\varPi TM$, the exterior derivative is represented
by the differential operator
\begin{align}
d=\sum_A dx^A \frac{\partial}{\partial x^A},
\end{align}
which acts on $C^\infty(\varPi TM)$ and is Grassmann odd.

Given a vector field
\begin{align}
X=\sum_A v^A(x)\frac{\partial}{\partial x^A},
\end{align}
one defines the contraction operator
\begin{align}
{\bm i}_X=\sum_A v^A(x)\frac{\partial}{\partial(dx^A)} ,
\end{align}
whose parity is
\begin{align}
|{\bm i}_X|=1-|X|.
\end{align}

The Lie derivative with respect to a vector field $X$
is defined geometrically by
\begin{align}
{\cal L}_X=[{\bm i}_X,\,d],
\end{align}
where the supercommutator is defined by
\begin{align}
[X,\,Y]=XY-(-1)^{|X||Y|}YX .
\end{align}
From this definition, we obtain the Cartan relations
\begin{align}
[d,\,{\cal L}_X] &=0,\\
[{\cal L}_X,\,{\bm i}_Y] &= {\bm i}_{[X,Y]},\\
[{\cal L}_X,\,{\cal L}_Y] &= {\cal L}_{[X,Y]}.
\end{align}
An explicit expression for ${\cal L}_X$ is
\begin{align}
{\cal L}_X
=X+
(-1)^{|X|}\sum_{A,B}dx^A
\frac{\partial v^B}{\partial x^A}\frac{\partial}{\partial(dx^B)} .
\end{align}

These relations constitute the Cartan calculus on supermanifolds,
which will be used in the following sections.

\subsection{Integral forms and picture-changing operators}

Unlike the purely bosonic case, differential forms on a supermanifold
do not admit a top form in the usual sense~\cite{Belopolsky:1996cy,Belopolsky:1997bg,
Belopolsky:1997jz,Witten:2012bg}.  The reason is that
even differentials such as $d\theta$ are not nilpotent, so ordinary
differential forms do not terminate at a maximal degree.  In order to
construct an integration theory, we introduce integral forms,
which contain delta functions of even differentials.

Formally, we introduce delta-function distributions $\delta(d\theta)$,
which are Grassmann odd.  For the vector field $\partial_\theta$, the
corresponding contraction operator is
\begin{align}
{\bm i}_{\partial_\theta}
=
\frac{\partial}{\partial(d\theta)} .
\end{align}
We then define the operator $\delta({\bm i}_{\partial_\theta})$ acting on a
form $\omega(d\theta)$ by
\begin{align}
\delta({\bm i}_{\partial_\theta})\,\omega(d\theta)
=
\int dt\,\omega(d\theta+t),
\end{align}
where $t$ is an even auxiliary parameter.
In particular, one has
\begin{align}
\delta({\bm i}_{\partial_\theta})\,\delta(d\theta)=1 .
\end{align}

More generally, for an odd vector field $X=\sum_A v^A(x)\frac{\partial}{\partial x^A}$
we define
\begin{align}
\delta({\bm i}_X)\,\omega(x,dx)
=
\int dt\,\omega(x,dx+t\,v).
\label{eq:deltaiX}
\end{align}
This operator is Grassmann odd.
Derivatives of the delta function are defined formally by
differentiation with respect to ${\bm i}_X$,
\begin{align}
\delta^{(n)}({\bm i}_X)
=
\frac{d^n}{d({\bm i}_X)^n}\delta({\bm i}_X).
\end{align}
These operators satisfy useful commutation relations such as
\begin{align}
[d,\,\delta^{(n)}({\bm i}_X)]
&=
\delta^{(n+1)}({\bm i}_X){\cal L}_X
+\frac{1}{2}\delta^{(n+2)}({\bm i}_X){\bm i}_{[X,\,X]} ,
\\
[{\cal L}_X,\,\delta^{(n)}({\bm i}_Y)]
&=
-\delta^{(n+1)}({\bm i}_Y){\bm i}_{[X,\,Y]} ,
\end{align}
for $|X|=|Y|=1$.

When $X$ is an odd vector field, we can define the operator
\begin{align}
\Gamma_X
=
\frac{1}{2}
\left\{\delta({\bm i}_X){\cal L}_X
-{\cal L}_X\delta({\bm i}_X)
\right\}.
\end{align}
This operator is Grassmann even and satisfies
\begin{align}
[d,\,\Gamma_X]=0.
\label{eq:dGammaX}
\end{align}
The operator $\Gamma_X$ plays a central role in the geometry of
integral forms and provides the geometric prototype of the
picture-changing operator in superstring theory\cite{Belopolsky:1997jz}.

Introducing a Grassmann parameter $\tau$,
the Lie derivative along an
odd vector field $X=\sum_Av^A(x)\frac{\partial}{\partial x^A}$ can be written as
\begin{align}
{\cal L}_X\omega(x,\,dx)
=
\int d\tau\,
\omega\!\left(
x+\tau v,\,
dx-\tau \sum_Adx^A\frac{\partial v}{\partial x^A}
\right).
\label{eq:calL_intexpression}
\end{align}
Using the representations (\ref{eq:deltaiX}) and (\ref{eq:calL_intexpression}),
one finds that $\Gamma_X$ admits the integral representation
\begin{align}
\Gamma_X\omega(x,\,dx)
=
\int {\cal D}(\tau,\,d\tau)\,
\omega(x+\tau v,d(x+\tau v)),
\label{eq:GammaXint}
\end{align}
where ${\cal D}(\tau,d\tau)$ denotes the measure over the auxiliary
parameters $(\tau,d\tau)$.
This expression is given by Eq.~(4.19) of \cite{Witten:2012bg}.

As a simple example of $\Gamma_X$ on $\varPi T \mathbb{C}^{1|1}$,
let us consider the odd vector field
\begin{align}
D=\frac{\partial}{\partial\theta}+\theta\frac{\partial}{\partial z} ,
\label{eq:Ddef}
\end{align}
which represents the supercovariant derivative on the
super Riemann surface and is Grassmann odd.  It anticommutes with the
supersymmetry generator
$Q=\frac{\partial}{\partial\theta}-\theta\frac{\partial}{\partial z}$.
Let us consider the following integral forms\cite{Belopolsky:1997jz}:
\begin{align}
\omega_1=\theta\,\delta(d\theta)\,f(z,\theta),
\qquad
\omega_2=dz\,\delta'(d\theta)\,f(z,\theta).
\end{align}
Using the definition of $\delta({\bm i}_D)$, one finds
\begin{align}
\delta({\bm i}_D)\omega_1&=-\theta f(z,\theta),
\qquad
\delta({\bm i}_D)\omega_2=+\theta f(z,\theta).
\end{align}
A straightforward computation then gives
\begin{align}
\Gamma_D\,\omega_1
&=
f(z,\theta)-\theta\frac{\partial f}{\partial\theta},
\qquad
\Gamma_D\,\omega_2
=
\theta\frac{\partial f}{\partial\theta}.
\end{align}
Combining these results, we obtain
\begin{align}
\Gamma_D\,
Y_D f(z,\theta)
=
f(z,\theta),
\qquad
Y_D = dz\,\delta^{\prime}(d\theta)+\theta\delta(d\theta).
\label{eq:GammaDidentity}
\end{align}
Thus, the integral-form factor 
$Y_D$
plays the role of an inverse picture-changing operator associated
with the odd vector field $D$,
characterized by the property that it is
removed by $\Gamma_D$\cite{Belopolsky:1997jz}.
This property will play an important role below. In particular,
the factor $Y_D$ will appear naturally in the descent
equations for vertex operators, 
while $\Gamma_D$ raises the picture number by one.

\section{Construction of NS-NS vertex operators from integral forms
via descent equations\label{sec:NSNSvertex}}

\subsection{Vertex operators in the NS-NS sector}

We consider superstring theory on a super Riemann surface
$\Sigma$ parametrized by local superconformal coordinates
\begin{align}
 (z,\theta,\bar{z},\bar{\theta}),
\end{align}
where $(z,\theta)$ and $(\bar{z},\bar{\theta})$ denote the holomorphic
and antiholomorphic coordinates, respectively.

A physical closed string state in the NS-NS sector is described by
a superfield vertex operator integrated over the full superspace:
\begin{align}
 \int d^2 z\, d^2\theta\,V(z,\theta,\bar{z},\bar{\theta}),
\label{eq:intV}
\end{align}
where $V(z,\theta,\bar{z},\bar{\theta})$ is a Grassmann even
superfield of conformal weight $(1/2,1/2)$.
It is constructed from matter superfields
\begin{align}
\bm{X}^\mu(z,\theta,\bar{z},\bar{\theta})
= X^\mu(z,\bar{z})
+ i\theta \psi^\mu(z)
+ i\bar{\theta}\tilde{\psi}^\mu(\bar{z}),
\end{align}
as in the standard formulation of superstring theory,
see, for example, Sec.~12.3 of \cite{Polchinski:1998rr}.

In the bosonic string, integrated and unintegrated vertex operators are
related via descent equations, as in \cite{Kishimoto:2024yuw}.
In the superstring case, however,
integration over odd directions requires integral forms with
delta-function insertions, leading to the notion of picture number.

Taking this structure into account,
we formulate vertex operators as integral forms
on the super Riemann surface.
Following \cite{Witten:2012bg}, integral forms are characterized
by their scaling behavior on $\varPi T\Sigma$ and by their localization
with respect to $d\theta$ and $d\bar{\theta}$.
A form with scaling weight $r$ that contains $n$ delta functions
of $d\theta$ and $d\bar{\theta}$
is said to have superdegree $m|n$, defined by
\begin{align}
m = r + n.
\end{align}
The exterior derivative increases the scaling weight by one
without changing $n$, and hence maps $m|n$-forms to
$m+1|n$-forms.
The corresponding picture number is $-n$.

The superspace integral \eqref{eq:intV} is naturally represented
by the following top integral form on $\Sigma$:
\begin{align}
 \omega_{2|2}
 = dz\,d\bar z\,
 \delta(d\theta)\,\delta(d\bar\theta)\,
 V(z,\theta,\bar z,\bar\theta).
\label{eq:NSNS-topform}
\end{align}
This is a top integral form on the 
super Riemann surface, with superdegree $2|2$ and
picture number $-2$.
The integrated vertex operator \eqref{eq:intV} is then expressed as
\begin{align}
 \int_{\Sigma}\omega_{2|2}
 = \int d^2z\, d^2\theta\,
 V(z,\theta,\bar z,\bar\theta).
\end{align}
Equivalently, this can be written as an integral over $\varPi T\Sigma$:
\begin{align}
 \int_{\Sigma}\omega_{2|2}
 =
 \int_{\varPi T\Sigma} {\cal D}(x,dx)\,
 \omega_{2|2}(x,dx),
\end{align}
where ${\cal D}(x,dx)$ is the natural Berezinian measure on $\varPi T\Sigma$
\cite{Witten:2012bg}.  Thus, the Berezin integration over the odd
coordinates is geometrically encoded by the delta functions
$\delta(d\theta)$ and $\delta(d\bar\theta)$.  Moreover, the top integral
form $\omega_{2|2}$ itself is superconformally invariant. 
This property is a direct consequence of the integral-form
formalism and will play a crucial role in the analysis
of descent equations below.

\subsection{The BRST transformation of $\omega_{2|2}$}

We now study the BRST transformation of the top integral form
$\omega_{2|2}$ defined in \eqref{eq:NSNS-topform}.
We denote its ghost number explicitly by a superscript,
writing $\omega_{2|2}^0$.

The BRST variation of $V(z,\theta,\bar{z},\bar{\theta})$
is determined by its transformation
as a superconformal primary field of weight $(1/2,1/2)$.
It is given by\footnote{
For a matter superconformal primary field $\Phi(z,\theta)$ of weight $h$,
the holomorphic part of the BRST transformation is given by
\begin{align*}
\delta_\mathrm{B} \Phi=
C \partial \Phi - \frac{1}{2}(D C) D \Phi + h (\partial C)\, \Phi .
\end{align*}
See, for example, \cite{Friedan:1985ge}.
The expression used in the text follows from this formula
by rewriting it in a total $D$-derivative form,
using the identity $D^2=\partial$.
}
\begin{align}
{\bm \delta}_\mathrm{B} V(z,\theta,\bar{z},\bar{\theta})
&=
D\Big\{\frac{1}{2}(DC)V - C DV \Big\}
+
\bar{D}\Big\{\frac{1}{2}(\bar{D}\tilde{C})V
- \tilde{C}\bar{D}V \Big\}.
\label{eq:BRST_V}
\end{align}
This shows that the BRST transformation
of the integrand (\ref{eq:intV})
is a total derivative on the super Riemann surface.
Here $C(z,\theta)$ and $\tilde{C}(\bar{z},\bar{\theta})$ are
the holomorphic and antiholomorphic ghost superfields.
The holomorphic ghost superfield is expanded as
\begin{align}
C(z,\theta) = c(z) + \theta\,\gamma(z),
\label{eq:C_to_c_gamma}
\end{align}
where $c(z)$ is the reparametrization ghost
and $\gamma(z)$ is its bosonic superpartner.
A similar expansion holds for $\tilde{C}(\bar{z},\bar{\theta})$.
The operator $D$ denotes the supercovariant derivative (\ref{eq:Ddef}),
and $\bar{D}$ is its antiholomorphic counterpart.

Here $dz, d\bar{z}, \delta(d\theta), \delta(d\bar{\theta})$
are Grassmann odd, while $d\theta, d\bar{\theta}$ are Grassmann even.
For delta functions of even variables,
we use the standard distributional identity
\begin{align}
 x\delta^{(n)}(x)=-n\delta^{(n-1)}(x),
\label{eq:delta-identity}
\end{align}
which follows by repeatedly differentiating
$x\delta(x)=0$.

Using this identity for the even differentials $d\theta$ and $d\bar{\theta}$,
we find that the BRST variation of $\omega_{2|2}^0$
takes the form
\begin{align}
{\bm \delta}_\mathrm{B} \omega_{2|2}^0
=
d \omega_{1|2}^1,
\end{align}
where $\omega_{1|2}^1$ is an integral form of superdegree $1|2$
and ghost number one given by
\begin{align}
\omega_{1|2}^1
&=
d\bar{z}\,\delta(d\bar{\theta})
Y_D
\left\{
\frac{1}{2}(DC)V - C DV
\right\}
+
dz\,\delta(d\theta)
Y_{\bar{D}}
\left\{
\frac{1}{2}(\bar{D}\tilde{C})V
- \tilde{C}\bar{D}V
\right\},
\label{eq:w1_1|2}
\end{align}
where $Y_D$ is defined in \eqref{eq:GammaDidentity}
in terms of the integral-form structure associated with $D$,
and $Y_{\bar{D}}$ is its antiholomorphic counterpart.

Thus, the BRST transformation of $\omega_{2|2}^0$
is exact with respect to the exterior derivative $d$.
This gives the first step in the descent equations,
relating a $2|2$ integral form to a $1|2$ one
while preserving the picture number.

\subsection{The BRST transformation of $\omega_{1|2}^1$
and the full set of descent equations}

We next consider the BRST transformation of the descendant
$\omega_{1|2}^1$ in (\ref{eq:w1_1|2}).
Using the BRST transformation \eqref{eq:BRST_V} of $V$ and
those of the ghost superfields,
\begin{align}
 \bm{\delta}_\mathrm{B}C
 &=
 C\partial C-\frac{1}{4}(DC)(DC),
 \qquad
 \bm{\delta}_\mathrm{B}\tilde{C}
 =
 \tilde{C}\bar{\partial}\tilde{C}
 -\frac{1}{4}(\bar{D}\tilde{C})(\bar{D}\tilde{C}),
\label{eq:BRST_C}
\end{align}
we find
\begin{align}
{\bm \delta}_\mathrm{B}\left\{
\frac{1}{2}(DC)V-CDV\right\}
&=
\bar{D}\hat{A},
\nonumber\\
{\bm \delta}_\mathrm{B}\left\{
\frac{1}{2}(\bar{D}\tilde{C})V-\tilde{C}\bar{D}V\right\}
&=
D\hat{A},
\label{eq:BRST-Ahat}
\end{align}
where
\begin{align}
\hat{A}
=
\left\{
\frac{1}{2}(DC)-CD
\right\}
\left\{
\frac{1}{2}(\bar{D}\tilde{C})
-\tilde{C}\bar{D}
\right\}V.
\label{eq:Ahat-def}
\end{align}
Substituting \eqref{eq:BRST-Ahat} into the BRST variation of
$\omega_{1|2}^1$, we obtain
\begin{align}
{\bm \delta}_{\mathrm B}\omega_{1|2}^1
&=
d\bar{z}\,\delta(d\bar{\theta})\,Y_D\bar{D}\hat{A}
+
dz\,\delta(d\theta)\,Y_{\bar{D}}D\hat{A}.
\end{align}
Using the explicit expressions of $Y_D$ and $Y_{\bar{D}}$,
together with (\ref{eq:delta-identity}),
a straightforward computation gives
\begin{align}
{\bm \delta}_{\mathrm B}\omega_{1|2}^1
=
d\omega_{0|2}^2,
\label{eq:second-descent}
\end{align}
where we define
\begin{align}
\omega_{0|2}^2
=
-Y_DY_{\bar{D}}\hat{A}
=
-Y_DY_{\bar{D}}
\left\{
\frac{1}{2}(DC)-CD
\right\}
\left\{
\frac{1}{2}(\bar{D}\tilde{C})
-\tilde{C}\bar{D}
\right\}V.
\label{eq:w2_0|2}
\end{align}
This is the second step of the descent equations at a fixed picture number.
We emphasize that the integral form $\omega_{0|2}^2$
defined in (\ref{eq:w2_0|2}) is superconformally invariant.

Finally, we examine the BRST transformation of $\omega_{0|2}^2$.
Since $Y_D$ and $Y_{\bar{D}}$ are BRST inert, it is enough
to evaluate the BRST variation of $\hat{A}$.
Using the BRST transformations \eqref{eq:BRST_V} and \eqref{eq:BRST_C},
one finds after a direct computation that
\begin{align}
{\bm \delta}_{\mathrm B}\hat{A}=0.
\end{align}
Therefore,
\begin{align}
{\bm \delta}_{\mathrm B}\omega_{0|2}^2=0.
\label{eq:w02-BRST-closed}
\end{align}

Collecting the above results, we obtain the descent equations
\begin{align}
{\bm \delta}_{\mathrm B}\omega_{2|2}^0
&=
d\omega_{1|2}^1,
\label{eq:deq22}\\
{\bm \delta}_{\mathrm B}\omega_{1|2}^1
&=
d\omega_{0|2}^2,
\label{eq:deq12}\\
{\bm \delta}_{\mathrm B}\omega_{0|2}^2
&=0.
\label{eq:deq02}
\end{align}
This completes the construction of the NS-NS vertex operators
with ghost numbers $0$, $1$, and $2$ at fixed picture number $-2$.\footnote{
In the present paper, we adopt the convention that all descent equations
take the form ${\bm \delta}_{\mathrm B}\omega = d\omega'$ without relative signs.
This differs from the convention used in \cite{Kishimoto:2024yuw},
where a relative sign appears in the first descent equation.
}

Having established the descent equations at a fixed picture number,
we now turn to their relation to the standard BRST formulation.
In particular, we will show that acting with the picture-changing
operators $\Gamma_D$ and $\Gamma_{\bar D}$ not only reproduces
the unintegrated vertex operator, but also reveals a fundamental
correspondence between supergeometric objects and ghost superfields.

\subsection{Picture-changing and the ghost-superfield correspondence
\label{eq:GammaD_dzC}}

In this subsection, we derive one of the main results of this paper:
a precise correspondence between supergeometric objects on
$\varPi T\Sigma$ and ghost superfields.
We now consider the action of the picture-changing operators
$\Gamma_D$ and $\Gamma_{\bar D}$ on the descent equations
constructed in the previous subsection.

Starting from the descent equations (\ref{eq:deq22}), (\ref{eq:deq12})
and (\ref{eq:deq02}),
we act with $\Gamma_D$ and $\Gamma_{\bar D}$,
which raise the holomorphic and antiholomorphic picture numbers,
respectively.
Since $\Gamma_D$ and $\Gamma_{\bar D}$ commute with the BRST operator
as well as with the exterior derivative $d$ (see (\ref{eq:dGammaX})),
they preserve the descent structure
and map BRST-closed forms to
BRST-closed ones.
Applying $\Gamma_D\Gamma_{\bar D}$ to the above sequence,
we obtain
\begin{align}
{\bm \delta}_{\mathrm B}\omega_{2|0}^0
&=
d\omega_{1|0}^1,
\label{eq:deq_dBomega_2|0}
\\
{\bm \delta}_{\mathrm B}\omega_{1|0}^1
&=
\,d\omega_{0|0}^2,
\\
{\bm \delta}_{\mathrm B}\omega_{0|0}^2
&=0,
\end{align}
where we have defined
\begin{align}
\omega_{m|0}^g
=
\Gamma_D\Gamma_{\bar D}\,\omega_{m|2}^g.
\end{align}

In particular, we obtain a BRST-invariant vertex operator
in picture number zero,
\begin{align}
\omega_{0|0}^2
=
\Gamma_D\Gamma_{\bar D}\,\omega_{0|2}^2.
\end{align}
This corresponds to the unintegrated vertex operator in the NS-NS sector.
From the expression (\ref{eq:w2_0|2}) and
the equations (\ref{eq:GammaDidentity})
and (\ref{eq:Ahat-def}), we finally obtain
\begin{align}
 \omega_{0|0}^2=
-\left\{\frac{1}{2}(DC)-CD\right\}
\left\{\frac{1}{2}(\bar{D}\tilde{C})-\tilde{C}\bar{D}\right\}V
\label{eq:w2_0|0}.
\end{align}
This provides a geometric derivation of the standard
unintegrated vertex operator from integral forms.

Here, the top form in (\ref{eq:deq_dBomega_2|0}) is rewritten as
\begin{align}
\omega_{2|0}^0=\left\{-d\theta -(dz-\theta d\theta)D\right\} 
\left\{-d\bar{\theta} -(d\bar{z}-\bar{\theta} d\bar{\theta})\bar{D}\right\}V.
\label{eq:w0_2|0}
\end{align}
Comparing this expression with (\ref{eq:w2_0|0}), we
can directly read off the correspondence
between the geometric differentials and the ghost superfields:
\begin{align}
dz-\theta d\theta  \leftrightarrow C,
\qquad
d\theta
\leftrightarrow
-\tfrac{1}{2}DC.
\label{eq:dzdthetaC}
\end{align}
This result reveals the geometric origin of the ghost superfield
and provides the central link between supergeometry and the BRST formulation.

We note that the one-form $dz-\theta d\theta$
coincides with the superconformal one-form $\varpi$
introduced in \cite{Witten:2012bg}.
Accordingly, the above relation can be rewritten as
$\varpi \leftrightarrow C$,
which shows that the BRST ghost superfield is nothing but
the geometric structure defining the super Riemann surface.
This relation reflects the fact that the BRST symmetry
originates from the gauge fixing of superdiffeomorphisms,
whose geometric structure is encoded in $\varpi$.

This correspondence can be understood more systematically
from infinitesimal supercoordinate transformations.
From Eq.~(42) in \cite{Friedan:1985ge},
the transformations generated by a super vector field
${\cal V}(z,\theta)$ are given by
\begin{align}
\delta z={\cal V}-\theta \tfrac{1}{2}D{\cal V},
\qquad
\delta\theta =\tfrac{1}{2}D{\cal V}.
\end{align}
Replacing ${\cal V}(z,\theta)$ with $\lambda\,C(z,\theta)$,
where $\lambda$ is an odd parameter,
we obtain
\begin{align}
\delta z =\lambda\left(C-\theta\tfrac{1}{2}DC\right),
\qquad
\delta\theta=\lambda\left(-\tfrac{1}{2}DC\right).
\end{align}
Interpreting these variations as differential forms,
we are naturally led to the identification
\begin{align}
dz \leftrightarrow C-\theta\tfrac{1}{2}DC,
\qquad
d\theta \leftrightarrow -\tfrac{1}{2}DC,
\end{align}
which is consistent with
$dz-\theta d\theta \leftrightarrow C$.

This derivation provides a geometric origin of the correspondence
between differential forms and ghost superfields.
In particular, the ghost superfield $C(z,\theta)$
can be regarded as the geometric counterpart of
the super one-form $dz-\theta d\theta$,
while its superderivative $DC$ corresponds to $d\theta$.
In this way, the action of $\Gamma_D$ and $\Gamma_{\bar D}$
translates the integral-form description into the standard BRST
formulation, establishing a direct bridge between
supergeometry and the operator formalism of superstring theory.

As a consistency check, let us examine the component
expansion of the superfield $V$:
\begin{align}
 V(z,\theta,\bar{z},\bar{\theta})
=V_{\theta\bar{\theta}}(z,\bar{z})+\theta V_{z\bar{\theta}}(z,\bar{z})
+\bar{\theta}V_{\theta\bar{z}}(z,\bar{z})
+\theta\bar{\theta}V_{z\bar{z}}(z,\bar{z}).
\end{align}
Substituting this into (\ref{eq:intV}), the integrated vertex operator is
reduced to
\begin{align}
 \int d^2 z\,d^2 \theta\,V(z,\theta,\bar{z},\bar{\theta})
=\int d^2z\,V_{z\bar{z}}(z,\bar{z}),
\end{align}
as expected.
We now turn to the component expansion of
$\omega_{0|0}^2$.
Substituting the above expansion of $V$ into
(\ref{eq:w2_0|0}), we obtain
\begin{align}
\omega_{0|0}^2
=&
-\frac{1}{4}\gamma\tilde{\gamma}V_{\theta\bar{\theta}}
+\frac{1}{2}\gamma\tilde{c}V_{\theta\bar{z}}
+\frac{1}{2}c\tilde{\gamma}V_{z\bar{\theta}}
+\tilde{c}cV_{z\bar{z}}
\nn
&+\theta\left\{
\frac{1}{4}\gamma\tilde{\gamma}V_{z\bar{\theta}}
-\frac{1}{2}\gamma\tilde{c}V_{z\bar{z}}
-\frac{1}{2}\tilde{\gamma}\left(\frac{1}{2}\partial c
+c\partial\right)\!
V_{\theta\bar{\theta}}
-\tilde{c}\left(
\frac{1}{2}\partial c+c\partial\right)\!
V_{\theta\bar{z}}
\right\}
\nn
&+\bar{\theta}\left\{
\frac{1}{4}\gamma\tilde{\gamma}V_{\theta\bar{z}}
+\frac{1}{2}c\tilde{\gamma}V_{z\bar{z}}
-\frac{1}{2}\gamma\left(\frac{1}{2}\bar{\partial}\tilde{c}
+\tilde{c}\bar{\partial}\right)\!
V_{\theta\bar{\theta}}
-c\left(\frac{1}{2}\bar{\partial}\tilde{c}
+\tilde{c}\bar{\partial}\right)\!
V_{z\bar{\theta}}\right\}
\nn
&+\theta\bar{\theta}
\left\{
-\frac{1}{4}\gamma\tilde{\gamma}V_{z\bar{z}}
+\left(\frac{1}{2}\partial c+c\partial\!\right)\!\left(
\frac{1}{2}\bar{\partial}\tilde{c}+\tilde{c}\bar{\partial}\!
\right)\!
V_{\theta{\bar{\theta}}}
-\frac{1}{2}\tilde{\gamma}\!\left(\frac{1}{2}\partial c+c\partial\!
\right)\! V_{\theta\bar{z}}
+\frac{1}{2}\gamma\!\left(\frac{1}{2}\bar{\partial}\tilde{c}+\tilde{c}\bar{\partial}\!
\right)\! 
V_{z\bar{\theta}}
\right\}.
\label{eq:component}
\end{align}
This reproduces the standard unintegrated vertex operator
$\tilde{c}c
V_{z\bar{z}}$ in the bosonic limit\footnote{
In the bosonic string case, the identification between the
two-form $d^2 z = -\frac{i}{2}\,dz \wedge d\bar{z}$ and the
ghost insertion leads naturally to a factor of $i$ in the
unintegrated vertex operator.
In the present analysis, we suppress this overall phase,
as it does not affect the structure of the descent equations.
}.
We also observe that
the terms involving $\theta$ or $\bar{\theta}$ are BRST exact:
\begin{align}
\omega_{0|0}^2
=&-\frac{1}{4}\gamma\tilde{\gamma}V_{\theta\bar{\theta}}
+\frac{1}{2}\gamma\tilde{c}V_{\theta\bar{z}}
+\frac{1}{2}c\tilde{\gamma}V_{z\bar{\theta}}
+\tilde{c}cV_{z\bar{z}}
\nn
&+\theta{\bm\delta}_{\rm B}\left(\tilde{c}V_{\theta\bar{z}}-\frac{1}{2}\tilde{\gamma} V_{\theta\bar{\theta}}\right)
+\bar{\theta}{\bm\delta}_{\rm B}\left(cV_{z\bar{\theta}}-\frac{1}{2}\gamma V_{\theta\bar{\theta}}\right)
\nn
&
+\theta\bar{\theta}{\bm\delta}_{\rm B}\left\{
-\frac{1}{2}\left(\frac{1}{2}\partial c+c\partial -\frac{1}{2}\bar{\partial}\tilde{c}-\tilde{c}\bar{\partial}\right)\! V_{\theta\bar{\theta}}
+\frac{1}{4}\gamma V_{z\bar{\theta}}-\frac{1}{4}\tilde{\gamma}V_{\theta\bar{z}}
\right\},
\end{align}
which shows that these terms do not contribute to the BRST cohomology.

\section{Vertex operators with Inverse picture-changing operators\label{sec:IPCO}}

In this section, we study vertex operators obtained by acting with inverse
picture-changing operators and derive the corresponding descent equations.
To avoid confusion, we do not explicitly indicate the worldsheet
superghost picture numbers in the notation of the operators.
Instead, the effect of the inverse picture-changing operators is encoded
in the symbols $\rho$, $\tilde{\rho}$, and $\mu$, corresponding to the
insertions of $Y$, $\tilde Y$, and $Y\tilde Y$, respectively.
Thus $\omega$, $\rho$, $\tilde{\rho}$, and $\mu$ correspond to
superghost picture numbers $(0,0)$, $(-1,0)$, $(0,-1)$, and $(-1,-1)$.

\subsection{Definition of $Y$ and $\tilde Y$}

We first introduce the inverse picture-changing operators associated with
the odd directions of the super Riemann surface.
Following Belopolsky~\cite{Belopolsky:1997jz},
the inverse picture-changing operator corresponding
to the odd vector field $D$ is written as
$Y_D$ in (\ref{eq:GammaDidentity}), which can be rewritten as
\begin{align}
Y_D=(dz -\theta d\theta)\delta^{\prime}(d\theta).
\label{eq:YDv2}
\end{align}
Under the identification (\ref{eq:dzdthetaC}),
this is represented as
\begin{align}
Y_D \;\longleftrightarrow\;
C\,\delta^{\prime}(-\tfrac{1}{2}DC).
\end{align}
Similarly, for the antiholomorphic sector, one obtains
\begin{align}
Y_{\bar{D}} \;\longleftrightarrow\;
\tilde{C}\,\delta^{\prime}(-\tfrac{1}{2}\bar{D}\tilde{C}).
\end{align}
Hence, 
for the superfield description, it is natural to introduce
\begin{align}
Y(z,\theta) \equiv C(z,\theta)\,\delta'(DC(z,\theta)),
\qquad
\tilde Y(\bar{z},\bar{\theta}) \equiv \tilde{C}(\bar{z},\bar{\theta})\,
\delta'(\bar{D}\tilde{C}(\bar{z},\bar{\theta})),
\end{align}
where $C(z,\theta)$ and $\tilde{C}(\bar{z},\bar{\theta})$ are
the holomorphic and antiholomorphic ghost superfields.
The delta functions are understood in the same sense as those
appearing in the integral form formalism.
In particular, although their arguments such as $DC$ (or $\gamma$)
are Grassmann even, the delta functions $\delta(DC)$ and their
derivatives are Grassmann odd.
It follows that the operator $Y$ is Grassmann even.

Expanding $Y(z,\theta)$ with (\ref{eq:C_to_c_gamma}), we obtain
\begin{align}
Y(z,\theta)
&=c\delta'(\gamma)-\theta\delta(\gamma)-\theta c(\partial c)\delta''(\gamma),
\end{align}
and it can be rewritten as
\begin{align}
 Y(z,\theta)&= c\delta'(\gamma)
+\theta \bm{\delta}_\mathrm{B}\{2c\delta''(\gamma)\},
\end{align}
where we used the distributional identity (\ref{eq:delta-identity}).
Therefore, 
it differs from the standard non-bosonized expression\footnote{
The non-bosonized expression of the inverse picture-changing operator
was given in \cite{Hata1987b,Yamron:1986nb}. A bosonized form
had already appeared
in \cite{Witten1986a},
where Witten notes that it had been constructed
by the authors of \cite{Friedan:1985ge} and described to him by E.~Martinec.}
$c\delta'(\gamma)$ by the BRST-exact term.\footnote{The BRST
transformations of the component fields, following from (\ref{eq:BRST_C}), are given by
\begin{align}
\bm{\delta}_\mathrm{B} c=c\partial c-\frac{1}{4}\gamma^2,
\quad
\bm{\delta}_\mathrm{B} \gamma=c\partial \gamma-\frac{1}{2}(\partial c)\gamma. 
\end{align}
}
An analogous relation holds for $\tilde Y$.

Although the superfield expressions contain these extra terms,
they are suitable for the descent construction below.
In particular, $Y$ and $\tilde Y$ are BRST closed:
\begin{align}
\bm{\delta}_\mathrm{B} Y =0,
\qquad
\bm{\delta}_\mathrm{B} \tilde Y =0.
\label{eq:BRST_Y}
\end{align}
For $Y$, this can be verified using (\ref{eq:BRST_C}) as follows:
\begin{align}
\bm{\delta}_\mathrm{B} Y
&=(\bm{\delta}_\mathrm{B} C)\delta'(DC)-C\,
\bm{\delta}_\mathrm{B}\!\left(\delta'(DC)\right)
\nn
&=
\left\{
C\partial C-\frac14(DC)(DC)
\right\}\delta'(DC)
+C\,
D\!\left\{
C\partial C-\frac14(DC)(DC)
\right\}
\delta''(DC)
=0,
\end{align}
and the proof for $\tilde Y$ is completely analogous.

It is also useful to note that $Y$ and $\tilde Y$ admit a compact
superfield representation, so that the action of inverse picture-changing
operators on superfields can be treated in a manifestly supersymmetric form.
In the following subsections, we apply these operators to the integrated
superfield vertex operator and derive the associated descent equations.

Another important property of $Y$ and $\tilde Y$ is their invariance under
superconformal transformations.\footnote{
We note that $Y_D$ and $Y_{\bar{D}}$ are invariant under superconformal transformations.
}
Let $(z,\theta)\to(z',\theta')$ be a superconformal transformation
(see, e.g.,  Sec.~12.3 in \cite{Polchinski:1998rr}).
The ghost superfield $C$ transforms as
\begin{align}
C'(z',\theta')=(D\theta')^2\,C(z,\theta),
\label{eq:superconformalC}
\end{align}
while the superderivative transforms as
\begin{align}
D'=(D\theta')^{-1}D.
\label{eq:superconformalD}
\end{align}
Using these transformation laws, we obtain
\begin{align}
D'C'
&=(D\theta')^{-1}D\left((D\theta')^2C\right)
=2C\,(\partial\theta')+(D\theta')\,DC.
\end{align}
Therefore,
\begin{align}
\delta'(D'C')
=\delta'\!\left((D\theta')DC\right)
+2C(\partial\theta')\,\delta''\!\left((D\theta')DC\right).
\end{align}
Using the scaling property of the delta function
in the sense of integral forms,
namely,  $\delta'(ax)=a^{-2}\delta'(x)$,
we find
\begin{align}
\delta'(D'C')
=(D\theta')^{-2}\delta'(DC)
+2C(\partial\theta')\,\delta''\!\left((D\theta')DC\right).
\end{align}
Combining this with the transformation (\ref{eq:superconformalC}), we obtain
\begin{align}
C'\delta'(D'C')
=(D\theta')^2C
\left\{
(D\theta')^{-2}\delta'(DC)
+2C(\partial\theta')\,\delta''\!\left((D\theta')DC\right)
\right\}.
\end{align}
The second term vanishes due to $C^2=0$ for
the ghost superfield, and we finally obtain
\begin{align}
C'\delta'(D'C')=C\delta'(DC).
\end{align}
Thus $Y=C\delta'(DC)$ is invariant under superconformal transformations.
The same argument applies to $\tilde Y$.

\subsection{Descent equations for vertex operators with inverse
picture-changing insertions}

We now apply the inverse picture-changing operators $Y$ and $\tilde Y$
to the top integral form $\omega_{2|2}^0$, whose explicit expression
was given in (\ref{eq:NSNS-topform}), and derive the associated
descent equations.

We introduce three new sequences of integral forms,
distinguished by the insertions of $Y$, $\tilde Y$, and $Y\tilde Y$:
\begin{align}
\rho_{2|2}^{-1} = Y\omega_{2|2}^0,
\qquad
\tilde{\rho}_{2|2}^{-1} = \tilde Y\omega_{2|2}^0,
\qquad
\mu_{2|2}^{-2} = Y\tilde Y\omega_{2|2}^0.
\label{eq:rhomu-topform}
\end{align}
The corresponding descendants are determined by requiring that
they satisfy descent equations of the form
\begin{align}
\bm{\delta}_\mathrm{B}\,\rho_{2|2}^{-1}
&=
d\rho_{1|2}^0,
\qquad
\bm{\delta}_\mathrm{B}\,\rho_{1|2}^0
=
d\rho_{0|2}^1,
\qquad
\bm{\delta}_\mathrm{B}\,\rho_{0|2}^1
=0,
\end{align}
and similarly for $\tilde{\rho}$ and $\mu$.
Here the superscript on $\rho$, $\tilde{\rho}$, and $\mu$
denotes the ghost number.
Since the inverse picture-changing operators $Y$ and $\tilde Y$
carry ghost number $-1$, the insertion of $Y$, $\tilde Y$,
or $Y\tilde Y$ shifts the ghost number by $-1$ or $-2$, respectively.

Since $Y$ and $\tilde Y$ are BRST closed (\ref{eq:BRST_Y}),
we obtain
\begin{align}
\bm{\delta}_\mathrm{B}\,\rho_{2|2}^{-1}
=
Yd\omega_{1|2}^1,
\qquad
\bm{\delta}_\mathrm{B}\,\tilde{\rho}_{2|2}^{-1}
=
\tilde{Y}d\omega_{1|2}^1,
\qquad
\bm{\delta}_\mathrm{B}\,\mu_{2|2}^{-2}
=
Y\tilde{Y}d\omega_{1|2}^1,
\end{align}
{}from (\ref{eq:deq22}).
At this stage, it is important to note that the exterior derivative
does not commute with multiplication by $Y$ or $\tilde Y$.
As a result, the descendants, $\rho_{1|2}^0$, $\tilde{\rho}_{1|2}^0$, and $\mu_{1|2}^{-1}$, 
are not simply obtained by acting
with $Y$ or $\tilde Y$ on $\omega_{1|2}^1$.
Instead, additional terms arise from the action of $d$ on the product
of $Y$ (or $\tilde Y$) with $\omega_{1|2}^1$.

Applying the BRST operator once more, we obtain
\begin{align}
\bm{\delta}_\mathrm{B}\rho_{1|2}^0
&=
d\rho_{0|2}^1,
\qquad
\bm{\delta}_\mathrm{B}\tilde{\rho}_{1|2}^0
=
d\tilde{\rho}_{0|2}^1,
\qquad
\bm{\delta}_\mathrm{B}\mu_{1|2}^{-1}
=
d\mu_{0|2}^0,
\end{align}
with
\begin{align}
\bm{\delta}_\mathrm{B}\,\rho_{0|2}^1
=
\bm{\delta}_\mathrm{B}\,\tilde{\rho}_{0|2}^1
=
\bm{\delta}_\mathrm{B}\,\mu_{0|2}^0
=0.
\end{align}
Interestingly, at the level of the lowest forms, the additional
terms drop out, and one recovers the simple relations
\begin{align}
\rho_{0|2}^1 = -Y \omega_{0|2}^2,
\qquad
\tilde{\rho}_{0|2}^1 = -\tilde{Y}\omega_{0|2}^2,
\qquad
\mu_{0|2}^0 = Y\tilde{Y}\omega_{0|2}^2.
\end{align}

Thus, the inverse picture-changing operators generate new
descent sequences starting from the top integral form.
While the lowest representatives are obtained by simple multiplication,
the intermediate descendants receive nontrivial additional contributions.
This structure will be derived explicitly below.

\subsection{Explicit derivation of the descent equations for $\rho$, $\tilde{\rho}$, and $\mu$}

In this subsection, we derive explicitly the descent equations
for the sequences $\rho$, $\tilde{\rho}$, and $\mu$
by direct BRST computation, and clarify the origin of the
additional terms appearing at intermediate levels.

\paragraph{The sequence $\rho$}

We start from
\begin{align}
\rho_{2|2}^{-1} = Y\omega_{2|2}^0,
\end{align}
where $\omega_{2|2}^0$
was given in (\ref{eq:NSNS-topform}).
Using the BRST invariance (\ref{eq:BRST_Y}) of $Y$,
we compute the BRST variation explicitly:
\begin{align}
\bm{\delta}_{\mathrm B}\rho_{2|2}^{-1}
=
dzd\bar{z}\,\delta(d\theta)\delta(d\bar{\theta})\,
Y\,\bm{\delta}_{\mathrm B}V.
\end{align}
Substituting \eqref{eq:BRST_V} and extracting total derivatives,
the expression $Y\bm{\delta}_{\mathrm B}V$ can be written as
a sum of total derivative terms together with an additional contribution:
\begin{align}
Y\bm{\delta}_\mathrm{B} V
=&D\left\{C\delta'(DC)\left(\frac{1}{2}(DC)V-CDV\right)\right\}
+\bar{D}\left\{C\delta'(DC)\left(\frac{1}{2}(\bar{D}\tilde{C})V
-\tilde{C}\bar{D}V\right)\right\}
\nn
&-D\{C\delta'(DC)\}\left\{\frac{1}{2}(DC)V-CDV\right\}.
\end{align}
The last term is expressed as
\begin{align}
-D\{C\delta'(DC)\}\left\{\frac{1}{2}(DC)V-CDV\right\}
=
D\left\{C\delta(DC)V\right\},
\label{eq:DCdeltaDC}
\end{align}
using the distributional identity (\ref{eq:delta-identity})
with $x=DC$.
Here, the combination $C\delta(DC)$
behaves as a superconformal primary field of weight $-1/2$,
reflecting its natural role in the descent construction
and the covariance of the resulting expressions.

As a result, we obtain
\begin{align}
\bm{\delta}_{\mathrm B}\rho_{2|2}^{-1}
=
d\rho_{1|2}^0,
\label{eq:tt4.28}
\end{align}
where
\begin{align}
\rho_{1|2}^0
&=
d\bar z\,\delta(d\bar\theta)Y_D
\frac{1}{2}
C\delta(DC)V
+
dz\,\delta(d\theta)Y_{\bar{D}}
Y\left\{
\frac{1}{2}(\bar D\tilde C)V-\tilde C\bar DV
\right\}.
\label{eq:rho1_1|2}
\end{align}
Here we have used
the expression $Y=C\delta'(DC)$ to organize the terms
in a compact form.

We next compute the BRST variation of $\rho_{1|2}^0$
and derive the second descent relation
$\bm{\delta}_{\mathrm B}\rho_{1|2}^0 = d\rho_{0|2}^1$.
In this computation, it is also important to use the BRST transformation
of the composite operator $C\delta(DC)$ together with
the distributional identities of delta functions.

We first compute the BRST variation of $C\delta(DC)$:
\begin{align}
\bm{\delta}_\mathrm{B}(C\delta(DC))
&=(\bm{\delta}_\mathrm{B}C)\delta(DC)-C\,\bm{\delta}_B\delta(DC)
\nonumber\\
&=\left\{
C\partial C-\frac{1}{4}(DC)(DC)\right\}\delta(DC)
+C\,D\left\{C\partial C-\frac{1}{4}(DC)(DC)\right\}\delta'(DC).
\end{align}
Using the identities of delta functions and their Grassmann oddness,
this simplifies to
\begin{align}
\bm{\delta}_\mathrm{B}\left\{C\delta(DC)\right\}
=\frac{1}{2}C \partial C \delta(DC).
\end{align}
Moreover, we have the following useful identities:
\begin{align}
 Y D\hat{A}
=&D\left\{
\frac{1}{2}C\delta(DC)
\left(
\frac{1}{2}(\bar{D}\tilde{C})-\tilde{C}\bar{D}\right)V\right\}
\end{align}
with (\ref{eq:Ahat-def})
and
\begin{align}
\bm{\delta}_\mathrm{B}(C\delta(DC)V)
=&\bar{D}\left\{
C\delta(DC)
\left(
\frac{1}{2}(\bar{D}\tilde{C})-\tilde{C}\bar{D}\right)V\right\}.
\end{align}
Using these identities above, we compute the BRST variation of $\rho_{1|2}^0$,
which leads to
\begin{align}
\bm{\delta}_\mathrm{B}\rho_{1|2}^0
=d\rho_{0|2}^1,
\label{eq:tt4.34}
\end{align}
where $\rho_{0|2}^1$ is
\begin{align}
\rho_{0|2}^1=&Y_DY_{\bar{D}}
\left\{
Y\left(\frac{1}{2}(DC)-CD\right)
\left(\frac{1}{2}(\bar{D}\tilde{C})-\tilde{C}\bar{D}\right)V
\right\}.
\end{align}

This result provides an explicit derivation of the relation
$\rho_{0|2}^1 = -Y\omega_{0|2}^2$
stated in the previous subsection.
Since both $Y$ and $\omega_{0|2}^2$ are BRST invariant,
we obtain
\begin{align}
 \bm{\delta}_\mathrm{B}\rho_{0|2}^1=0,
 \label{eq:tt4.36}
\end{align}
which completes the descent equations for the $\rho$ sequence.

In the same way as for the sequence $\omega$,
we can construct a picture-zero sequence from $\rho$
by acting with the picture-changing operators
$\Gamma_D$ and $\Gamma_{\bar D}$.
Since $\Gamma_D$ and $\Gamma_{\bar D}$
commute with the BRST operator, they preserve the descent structure.
Applying $\Gamma_D\Gamma_{\bar D}$ to the sequence
$\rho_{r|2}^g$ in (\ref{eq:tt4.28}), (\ref{eq:tt4.34}), and (\ref{eq:tt4.36}), we obtain
\begin{align}
\bm{\delta}_\mathrm{B}\rho_{2|0}^{-1}
=
d\rho_{1|0}^0,
\qquad
\bm{\delta}_\mathrm{B}\rho_{1|0}^0
=
d\rho_{0|0}^1,
\qquad
\bm{\delta}_\mathrm{B}\rho_{0|0}^1
=0,
\end{align}
where we define
\begin{align}
\rho_{m|0}^g
=
\Gamma_D\Gamma_{\bar D}\,\rho_{m|2}^g.
\end{align}
In particular, we obtain the BRST-invariant unintegrated vertex operator
\begin{align}
\rho_{0|0}^1
=
\Gamma_D\Gamma_{\bar D}\,\rho_{0|2}^1.
\end{align}
Using the relation $\rho_{0|2}^1=-Y\omega_{0|2}^2$
and the action of $\Gamma_D\Gamma_{\bar D}$,
this can be written as
\begin{align}
\rho_{0|0}^1
=
-\,Y\omega_{0|0}^2.
\end{align}
Thus $\rho_{0|0}^1$ is obtained from the standard unintegrated
NS-NS vertex operator by a single inverse picture-changing insertion.

\paragraph{The sequence $\tilde{\rho}$}

The analysis for the sequence $\tilde{\rho}$ proceeds in exactly the same way
as for $\rho$.
The computation is obtained by exchanging the holomorphic and
antiholomorphic sectors.
We therefore omit the details since the computation is identical
and summarize the result.

The descent equations take the form
\begin{align}
\bm{\delta}_\mathrm{B}\,\tilde{\rho}_{2|2}^{-1}
=
d\tilde{\rho}_{1|2}^{0},
\qquad
\bm{\delta}_\mathrm{B}\,\tilde{\rho}_{1|2}^{0}
=
\,d\tilde{\rho}_{0|2}^1,
\qquad
\bm{\delta}_\mathrm{B}\,\tilde{\rho}_{0|2}^1
=0,
\end{align}
where
\begin{align}
\tilde{\rho}_{1|2}^{-1}
=&d\bar{z}\delta(d\bar{\theta})Y_D\left\{
\tilde{Y}\left(
\frac{1}{2}(DC)V-CDV\right)\right\}
+dz\delta(d\theta)Y_{\bar{D}}
\frac{1}{2}\tilde{C}\delta(\bar{D}\tilde{C})V,
\\
\tilde{\rho}_{0|2}^1
=&Y_DY_{\bar{D}}
\left\{
\tilde{Y}\left(\frac{1}{2}(DC)-CD\right)
\left(\frac{1}{2}(\bar{D}\tilde{C})-\tilde{C}\bar{D}\right)V\right\}
=
-\,\tilde{Y}\omega_{0|2}^2.
\end{align}
Applying the picture-changing operators,
we obtain the picture-zero sequence
\begin{align}
\bm{\delta}_\mathrm{B}\tilde{\rho}_{2|0}^{-1}
=
d\tilde{\rho}_{1|0}^0,
\qquad
\bm{\delta}_\mathrm{B}\tilde{\rho}_{1|0}^0
=
d\tilde{\rho}_{0|0}^1,
\qquad
\bm{\delta}_\mathrm{B}\tilde{\rho}_{0|0}^1
&=0,
\end{align}
with
\begin{align}
\tilde{\rho}_{r|0}^g
=
\Gamma_D\Gamma_{\bar D}\,\tilde{\rho}_{r|2}^g.
\end{align}
In particular, the unintegrated vertex operator is given by
$\tilde{\rho}_{0|0}^1=-\tilde{Y}\omega_{0|0}^2$.

\paragraph{The sequence $\mu$}

The derivation for the sequence $\mu$ proceeds similarly to those for
$\rho$ and $\tilde{\rho}$, now involving both inverse picture-changing
operators $Y$ and $\tilde{Y}$.
Using the BRST invariance of $Y$ and $\tilde{Y}$ and the same
distributional identities of delta functions as before,
we obtain the descent equations
\begin{align}
\bm{\delta}_\mathrm{B}\mu_{2|2}^{-2}
=
d\mu_{1|2}^{-1},
\qquad
\bm{\delta}_\mathrm{B}\mu_{1|2}^{-1}
=
\,d\mu_{0|2}^0,
\qquad
\bm{\delta}_\mathrm{B}\mu_{0|2}^0
=0.
\end{align}
Here, $\mu_{1|2}^1$ and $\mu_{0|2}^2$ are defined by
\begin{align}
\mu_{1|2}^{-1}=&d\bar{z}\delta(d\bar{\theta})Y_D
\frac{1}{2}\tilde{Y}C\delta(DC)V
+dz\delta(d\theta)Y_{\bar{D}}
\frac{1}{2}Y\tilde{C}\delta(\bar{D}\tilde{C})V,
\\
\mu_{0|2}^0=&-Y_DY_{\bar{D}}Y\tilde{Y}
\left\{\frac{1}{2}(DC)-CD\right\}\left\{\frac{1}{2}(\bar{D}\tilde{C})
-\tilde{C}\bar{D}
\right\}V=Y\tilde{Y}\omega_{0|2}^2.
\end{align}

Applying the picture-changing operators $\Gamma_D$ and $\Gamma_{\bar D}$,
we obtain the picture-zero sequence
\begin{align}
\bm{\delta}_\mathrm{B}\mu_{2|0}^{-2}
=
\,d\mu_{1|0}^{-1},
\qquad
\bm{\delta}_\mathrm{B}\mu_{1|0}^{-1}
=
\,d\mu_{0|0}^0,
\qquad
\bm{\delta}_\mathrm{B}\mu_{0|0}^0
=0,
\end{align}
with
\begin{align}
\mu_{m|0}^g
=
\Gamma_D\Gamma_{\bar D}\,\mu_{m|2}^g.
\end{align}
In particular, the unintegrated vertex operator is given by
\begin{align}
\mu_{0|0}^0
=
\Gamma_D\Gamma_{\bar D}\,\mu_{0|2}^0
=
Y\tilde{Y}\omega_{0|0}^2.
\end{align}

We note that
$\mu_{0|0}^0$ admits a particularly simple expression:
\begin{align}
\mu_{0|0}^0=&-\frac{1}{4}C\delta(DC)\tilde{C}\delta(\bar{D}\tilde{C})V.
\end{align}
Expanding this in component fields, we obtain
\begin{align}
\mu_{0|0}^0=& \frac{1}{4}c\tilde{c}\delta(\gamma)\delta(\tilde{\gamma})
V_{\theta\bar{\theta}}
+\cdots,
\end{align}
where the ellipsis denotes terms involving $\theta$ or
$\bar{\theta}$, which are BRST exact.
This expression makes it clear that $\mu_{0|0}^0$
contains the usual $(-1,-1)$-picture vertex operator.

\section{Higher ghost number vertex operators\label{sec:gh3}}

In the bosonic string, vertex operators with higher ghost number
can be constructed by multiplying the operator
by the factor $(\partial c-\bar{\partial}\tilde c)$,
which generates the next element in the descent sequence
\cite{Kishimoto:2024yuw}.

In this section, we extend this construction to the superstring.
Our starting point is the natural superfield analogue
of $(\partial c-\bar{\partial}\tilde c)$:
\begin{align}
\partial C(z,\theta)-\bar{\partial}\tilde C(\bar{z},\bar{\theta}),
\end{align}
where $C(z,\theta)$ and $\tilde{C}(\bar{z},\bar{\theta})$ are the ghost
superfields\footnote{
The choice of the antisymmetric combination 
$\partial C-\bar{\partial}\tilde{C}$ is motivated by the expectation
that the symmetric combination is BRST-trivial in the absolute cohomology,
as suggested in the literature (see section~2.3 in \cite{Thorn1989}),
although a complete proof is not known.}.  
We investigate whether vertex operators obtained by
multiplying the NS-NS vertex operator by this factor satisfy descent
equations in the BRST formalism.

\subsection{NS-NS vertex operators with $\partial C-\bar{\partial}\tilde C$}

We introduce top-form operators obtained by multiplying
the NS-NS vertex operators in (\ref{eq:NSNS-topform})
and (\ref{eq:rhomu-topform}) by the factor
$\partial C-\bar{\partial}\tilde{C}$ :
\begin{align}
\omega_{2|2}^1=&
dzd\bar{z}\delta(d\theta)\delta(d\bar{\theta})
(\partial C-\bar{\partial}\tilde{C})V,
\label{eq:w_2|2^1}\\
\rho_{2|2}^0=&
dzd\bar{z}\delta(d\theta)\delta(d\bar{\theta})
(\partial C-\bar{\partial}\tilde{C})YV,
\label{eq:rho_2|2^0}
\\
\tilde{\rho}_{2|2}^0=&
dzd\bar{z}\delta(d\theta)\delta(d\bar{\theta})
(\partial C-\bar{\partial}\tilde{C})\tilde{Y}V,
\label{eq:tilderho_2|2^0}
\\
\mu_{2|2}^{-1}=&
dzd\bar{z}\delta(d\theta)\delta(d\bar{\theta})
(\partial C-\bar{\partial}\tilde{C})Y\tilde{Y}V.
\label{eq:mu_2|2^-1}
\end{align}
These operators carry one additional unit of ghost number 
compared to  (\ref{eq:NSNS-topform}) and (\ref{eq:rhomu-topform}).

First, we compute the BRST variation of $(\partial C-\bar{\partial}\tilde{C})V$ 
simply by using (\ref{eq:BRST_V}) and (\ref{eq:BRST_C}):
\begin{align}
\bm{\delta}_\mathrm{B}
\left\{
(\partial C-\bar{\partial}\tilde{C})V
\right\}
=&D\left\{
(\partial C-\bar{\partial}\tilde{C})\left(\frac{1}{2}(DC)V-CDV\right)-C(D^3C)V
\right\}
\nn
&+\bar{D}\left\{
(\partial C-\bar{\partial}\tilde{C})\left(
\frac{1}{2}(\bar{D}\tilde{C})V-\tilde{C}\bar{D}V
\right)+\tilde{C}(\bar{D}^3\tilde{C})V\right\}.
\end{align}
This structure shows that the BRST variation is expressed
as a sum of superderivatives, which directly leads to
the exterior derivative acting on the corresponding integral forms.
From this expression, we find that the descent equation is satisfied:
\begin{align}
 \bm{\delta}_\mathrm{B}\omega_{2|2}^1=&
d\omega_{1|2}^2,
\end{align}
where $\omega_{1|2}^2$ is defined as
\begin{align}
\omega_{1|2}^2=&dz\delta(d\theta)Y_{\bar{D}}\left\{
(\partial C-\bar{\partial}\tilde{C})\left(
\frac{1}{2}(\bar{D}\tilde{C})V-\tilde{C}\bar{D}V
\right)
+\tilde{C}(\bar{D}^3\tilde{C})V\right\}\nn
&+d\bar{z}\delta(d\bar{\theta})Y_D
\left\{
(\partial C-\bar{\partial}\tilde{C})\left(\frac{1}{2}(DC)V
-CDV\right)-C(D^3C)V
\right\}.
\label{eq:w_1|2^2}
\end{align}

It is important to emphasize that $\omega_{1|2}^2$
is not simply given by multiplying
$(\partial C-\bar{\partial}\tilde{C})$
with the corresponding lower-form operator.
In contrast to the bosonic case, additional terms of the form
$C(D^3C)V$ and $\tilde{C}(\bar{D}^3\tilde{C})V$
necessarily appear.
These contributions play a crucial role in the consistency
of the descent equations, as they provide the necessary
cancellations in the subsequent BRST transformation.

A straightforward but lengthy computation leads to the next step of the
descent equations.
The result is given by
\begin{align}
 \bm{\delta}_\mathrm{B}\omega_{1|2}^2=&d\omega_{0|2}^3,
\\
 \bm{\delta}_\mathrm{B}\omega_{0|2}^3=&0,
\end{align}
where $\omega_{0|2}^3$ takes the explicit form:
\begin{align}
\omega_{0|2}^3=&\;
-Y_DY_{\bar{D}}\left[
(\partial C-\bar{\partial}\tilde{C})
\left\{\frac{1}{2}(\bar{D}\tilde{C})-\tilde{C}\bar{D}
\right\}
\left\{
\frac{1}{2}(DC)-CD
\right\}V\right.
\nn
&
\quad
\left.+\tilde{C}(\bar{D}^3\tilde{C})
\left\{
\frac{1}{2}(DC)-CD\right\}V
-C(D^3C)\left\{
\frac{1}{2}(\bar{D}\tilde{C})-\tilde{C}\bar{D}
\right\}V\right]
\label{eq:n0}.
\end{align}

The construction presented above for the $\omega$ sequence
extends straightforwardly to all picture sectors (\ref{eq:rho_2|2^0}),
(\ref{eq:tilderho_2|2^0}) and (\ref{eq:mu_2|2^-1}).
The derivation proceeds in complete analogy with that of the $\omega$ sequence.
In particular, the BRST invariance of $Y$ and $\tilde{Y}$ ensures that
the same structure of descent equations is preserved.

The first descent equation is given by
\begin{align}
 \bm{\delta}_\mathrm{B}\rho_{2|2}^0=d\rho_{1|2}^1,
\quad
 \bm{\delta}_\mathrm{B}\tilde{\rho}_{2|2}^0=d\tilde{\rho}_{1|2}^1,
\quad
 \bm{\delta}_\mathrm{B}\mu_{2|2}^{-1}=d\mu_{1|2}^0,
\end{align}
where explicit expressions for the lower integral forms are given as follows:
\begin{align}
\rho_{1|2}^1
=&
dz\delta(d\theta)Y_{\bar{D}}
\left[
(\partial C-\bar{\partial}\tilde{C})
\left\{
\frac{1}{2}(\bar{D}\tilde{C})-\tilde{C}\bar{D}
\right\}YV
+\tilde{C}(\bar{D}^3\tilde{C})YV
\right]
\nn
&+
d\bar{z}\,\delta(d\bar{\theta})Y_D
\left\{
(\partial C-\bar{\partial}\tilde{C})
\frac{1}{2}C\delta(DC)V
\right\},
\label{eq:rho_1|2^1}
\\
\tilde{\rho}_{1|2}^1
=&
dz\delta(d\theta)Y_{\bar{D}}
\left\{
(\partial C-\bar{\partial}\tilde{C})
\frac{1}{2}\tilde{C}\delta(\bar{D}\tilde{C})V
\right\}
\nn
&+
d\bar{z}\,\delta(d\bar{\theta})Y_D
\left[
(\partial C-\bar{\partial}\tilde{C})
\left\{
\frac{1}{2}(DC)
- C D
\right\}\tilde{Y}V
- C(D^3C)\tilde{Y}V
\right],
\label{eq:tilderho_1|2^1}
\\
\mu_{1|2}^0
=&
dz\delta(d\theta)Y_{\bar{D}}
\left\{
(\partial C-\bar{\partial}\tilde{C})
\frac{1}{2}\tilde{C}\delta(\bar{D}\tilde{C})YV
\right\}
+d\bar{z}\delta(d\bar{\theta})Y_D
\left\{
(\partial C-\bar{\partial}\tilde{C})
\frac{1}{2}C\delta(DC)\tilde{Y}V
\right\}.
\label{eq:mu_1|2^0}
\end{align}
Similarly, the remaining equations are given by
\begin{align}
&
 \bm{\delta}_\mathrm{B}\rho_{1|2}^1=d\rho_{0|2}^2,
\quad
 \bm{\delta}_\mathrm{B}\tilde{\rho}_{1|2}^1=d\tilde{\rho}_{0|2}^2,
\quad
 \bm{\delta}_\mathrm{B}\mu_{1|2}^0=d\mu_{0|2}^1,
\nn
&
 \bm{\delta}_\mathrm{B}\rho_{0|2}^2=0,
\quad
 \bm{\delta}_\mathrm{B}\tilde{\rho}_{0|2}^2=0,
\quad
 \bm{\delta}_\mathrm{B}\mu_{0|2}^1=0,
\end{align}
where the forms are
\begin{align}
\rho_{0|2}^2
=&\;
-Y_DY_{\bar{D}}
\left[
(\partial C-\bar{\partial}\tilde{C})
\frac{1}{2}C\delta(DC)
\left\{
\frac{1}{2}(\bar{D}\tilde{C})-\tilde{C}\bar{D}
\right\}V
\right.
\nn
&\quad
\left.
+\tilde{C}(\bar{D}^3\tilde{C})
\frac{1}{2}C\delta(DC)V
\right],
\\
\tilde{\rho}_{0|2}^2
=&\;
-Y_DY_{\bar{D}}
\left[
(\partial C-\bar{\partial}\tilde{C})
\frac{1}{2}\tilde{C}\delta(\bar{D}\tilde{C})
\left\{
\frac{1}{2}(DC)-CD
\right\}V
\right.
\nn
&\quad
\left.
- C(D^3C)
\frac{1}{2}\tilde{C}\delta(\bar{D}\tilde{C})V
\right],
\\
\mu_{0|2}^1
=&
-Y_DY_{\bar{D}}
(\partial C-\bar{\partial}\tilde{C})
\,\frac{1}{4}
C\delta(DC)\,
\tilde{C}\delta(\bar{D}\tilde{C})\,V.
\label{eq:mu_0|2^1}
\end{align}

Following the general prescription established above,
we construct the corresponding unintegrated vertex operator
by the action of $\Gamma_D\Gamma_{\bar D}$:
\begin{align}
\omega_{0|0}^3
&=
\Gamma_D\Gamma_{\bar D}\,\omega_{0|2}^3,
\\
\rho_{0|0}^2
&=
\Gamma_D\Gamma_{\bar D}\,\rho_{0|2}^2,
\\
\tilde{\rho}_{0|0}^2
&=
\Gamma_D\Gamma_{\bar D}\,\tilde{\rho}_{0|2}^2,
\\
\mu_{0|0}^1
&=
\Gamma_D\Gamma_{\bar D}\,\mu_{0|2}^1.
\end{align}
This provides a unified construction of unintegrated
vertex operators across all picture sectors.

\subsection{Superconformal properties of the vertex operators}

We analyze the superconformal transformation properties
of the vertex operators constructed in the previous subsection.
Without the factor $(\partial C-\bar{\partial}\tilde{C})$,
their invariance is manifest.
In its presence, however, the invariance becomes nontrivial
and requires explicit verification.

The non-invariant terms are expressed as
$\bm{\delta}_{\mathrm B}$-exact contributions for the $0|2$-forms,
while for higher forms they appear as a sum of
$\bm{\delta}_{\mathrm B}$-exact and $d$-exact terms,
so that the descent structure is preserved under
superconformal transformations.
This structure is analogous to that in
the bosonic string \cite{Kishimoto:2024yuw}.

To analyze this structure, we examine the transformation properties
of the basic superfields and superderivatives,
as well as those of the composite operators appearing
in the vertex operators.
The additional terms in the descent equations
play an essential role in ensuring the correct transformation properties.

We first recall that the inverse picture-changing operators
$Y$ and $\tilde{Y}$ are invariant under superconformal transformations.
For instance, under a superconformal transformation
$(z,\theta)\to(z',\theta')$, the ghost superfield $C$ and
the superderivative $D$ transform as (\ref{eq:superconformalC})
and (\ref{eq:superconformalD}).
Using the transformation of $DC$ and the scaling property of delta functions,
we find that $Y=C\delta'(DC)$ is invariant.
Moreover, as noted below (\ref{eq:DCdeltaDC}),
both $C\delta(DC)$ and $C\partial C\,\delta(DC)$
transform as superconformal primary fields of weight $-1/2$.
This follows from the transformation of $\partial C$,
\begin{align}
 \partial C=\partial' C'+(D\theta')^{-2}(\partial \theta')D'C'
-2(D\theta')^{-3}(D^3\theta')C'.
\label{eq:superconformal_dC}
\end{align}
The same holds for $\tilde{Y}$.
Using these properties,
the extra terms are annihilated by
$\delta(DC)$ and hence
$\mu_{0|2}^1$ is invariant.

We next examine the behavior of the integral forms involving
$\partial C-\bar{\partial}\tilde{C}$
under superconformal transformations.
In contrast to the invariance of $\mu_{0|2}^1$,
the remaining operators
are not strictly invariant.
To make this structure manifest, let us introduce
\begin{align}
{\mathcal C}^{(2)}
&=
\partial C\left(\frac12\,DC-CD\right)-C(D^3C)
=
D(\bm{\delta}_{\mathrm B}C)+C\,\partial C\,D,
\label{eq:star-def}
\\
\tilde{\mathcal C}^{(2)}
&=
\bar{\partial}\tilde C
\left(\frac12\,\bar D\tilde C-\tilde C\bar D\right)
-\tilde C(\bar D^3\tilde C)
=
\bar D(\bm{\delta}_{\mathrm B}\tilde C)
+\tilde C\,\bar{\partial}\tilde C\,\bar D .
\label{eq:tildestar-def}
\end{align}
Using these quantities, the operators
can be rewritten as
\begin{align}
\omega_{0|2}^{3}
=&
-Y_DY_{\bar{D}}
\Biggl[
\mathcal{C}^{(2)}
\left(\frac{1}{2}\,\bar D\tilde C-\tilde C\bar D\right)V
-
\tilde{\mathcal C}^{(2)}
\left(\frac{1}{2}\,DC-CD\right)V
\Biggr],
\label{eq:w_0|2^3-rewrite}
\\
\rho_{0|2}^2
=&
-Y_DY_{\bar{D}}
\Biggl[
-\frac{1}{2}
\tilde{\mathcal C}^{(2)}
C\,\delta(DC)\,V
-\frac12 C\,\partial C\,\delta(DC)
\left(\frac{1}{2}\,\bar D\tilde C-\tilde C\bar D\right)V
\Biggr],
\label{eq:tilderho_0|2-rewrite}
\\
\tilde{\rho}_{0|2}^2
=&
-Y_DY_{\bar{D}}
\Biggl[
\frac{1}{2}
{\mathcal C}^{(2)}
\tilde C\,\delta(\bar D\tilde C)\,V
+
\frac{1}{2}\tilde C\,\bar{\partial}\tilde C\,
\delta(\bar D\tilde C)
\left(\frac{1}{2}\,DC-CD\right)V
\Biggr].
\label{eq:rho_0|2^2-rewrite}
\end{align}

The crucial point is that 
${\mathcal C}^{(2)}$
 does not transform as a primary by
itself. Under a superconformal transformation
$(z,\theta)\to(z',\theta')$, one finds
\begin{align}
{\mathcal C}^{(2)}
(D\theta^{\prime})
=
{\mathcal C}^{(2)\prime}
-
(D\theta')^{-2}(\partial\theta')
\left(
C'\partial' C'
-\frac{1}{2} (D'C')^2
+
C'(D'C')D'
\right).
\label{eq:star-transform}
\end{align}
Combining this with the transformation laws of the other factors, one
obtains
\begin{align}
{\mathcal C}^{(2)}
\frac{1}{2}\tilde C\,\delta(\bar D\tilde C)\,V
=&
{\mathcal C}^{(2)\prime}
\frac{1}{2}\tilde C'\,\delta(\bar D'\tilde C')\,V'
-
(D\theta')^{-2}(\partial\theta')\,
\bm{\delta}_{\mathrm B}
\!\left[
C'\tilde C'\,\delta(\bar D'\tilde C')\,V'
\right],
\label{eq:star-term-transform-1}
\\
{\mathcal C}^{(2)}
\left(\frac12\,\bar D\tilde C-\tilde C\bar D\right)V
=&
{\mathcal C}^{(2)\prime}
\left(\frac12\,\bar D'\tilde C'-\tilde C'\bar D'\right)V'
-
(D\theta')^{-2}(\partial\theta')\,
\bm{\delta}_{\mathrm B}
\!\left[
2C'\left(\frac12\,\bar D'\tilde C'-\tilde C'\bar D'\right)V'
\right].
\label{eq:star-term-transform-2}
\end{align}
Thus, the non-primary behavior of 
${\mathcal C}^{(2)}$ is entirely captured
by a $\bm{\delta}_\mathrm{B}$-exact term.
The corresponding formulas in the antiholomorphic sector are obtained
by exchanging the holomorphic and antiholomorphic variables.

Collecting the holomorphic and antiholomorphic contributions, we arrive
at
\begin{align}
\omega_{0|2}^{3}
=&
\omega_{0|2}^{3\,\prime}
+
\bm{\delta}_{\mathrm B}\psi_{0|2}^{2\,(0,0)},
\\
\rho_{0|2}^{2}
=&
\rho_{0|2}^{2\,\prime}
+
\bm{\delta}_{\mathrm B}\psi_{0|2}^{1\,(-1,0)},
\\
\bar{\rho}_{0|2}^{2}
=&
\bar{\rho}_{0|2}^{2\,\prime}
+
\bm{\delta}_{\mathrm B}\psi_{0|2}^{1\,(0,-1)},
\end{align}
where we have defined as
\begin{align}
\psi_{0|2}^{2\,(0,0)}
=&
-Y_DY_{\bar{D}}
\Bigl[
(D\theta')^{-2}(\partial\theta')\,C'
\bigl(\bar D'\tilde C'-2\tilde C'\bar D'\bigr)V'
\nonumber\\
&\qquad
-
(\bar D\bar\theta')^{-2}(\bar\partial\bar\theta')\,\tilde C'
\bigl(D'C'-2C'D'\bigr)V'
\Bigr],
\\ 
\psi_{0|2}^{1\,(-1,0)}
=&
Y_DY_{\bar{D}}
(\bar D\bar\theta')^{-2}(\bar\partial\bar\theta')
\tilde C'C'\delta(D'C')\,V',
\\
\psi_{0|2}^{1\,(0,-1)}
=&
-Y_DY_{\bar{D}}
(D\theta')^{-2}(\partial\theta')
C'\tilde C'\delta(\bar D'\tilde C')\,V'.
\end{align}
We denote by $\psi_{m|n}^{g,(p,\bar{p})}$ the operators
that appear in the non-invariant part of the transformation laws.
These operators do not form an independent descent sequence,
but appear as compensating terms that restore the covariance
of the descent equations under superconformal transformations.

Therefore, although
the $0|2$-forms in the  $(0,0),\,(-1,0)$ and $(0,-1)$ pictures
are not strictly superconformally invariant as local integral forms,
their non-invariant parts are $\bm{\delta}_{\mathrm B}$-exact.
Therefore, they define the same BRST cohomology class
and are superconformally well-defined at the cohomological level,
even though they are not strictly invariant as local operators.

We next examine the transformation properties of the $1|2$-forms
appearing in the descent sequence.
In contrast to the $0|2$-forms, their transformation laws involve
both $\bm{\delta}_{\mathrm B}$-exact and total-derivative terms.

First, we can find the useful equations:
\begin{align}
 \partial C Y (D\theta')
=&
(D\theta')
\left[
\partial' C' Y'-\left(
D'\log\frac{D\theta'}{\bar{D}\bar{\theta}'}\right)
C'\delta(D'C')
\right]
\end{align}
and
\begin{align}
 \bm{\delta}_\mathrm{B}
\left[
Y\tilde{C}\delta(\bar{D}\tilde{C})V\right]
=&
\frac{1}{2}D\left[
C\delta(DC)\tilde{C}\delta(\bar{D}\tilde{C})V\right].
\end{align}
Noting that
\begin{align}
 Y_D\left(dz\partial+d\theta\frac{\partial}{\partial \theta}\right)
=-dz\delta(d\theta)D,
\quad
dY_DY_{\bar{D}}
=&
-d\bar{z}\delta(d\bar{\theta})Y_D
\bar{D}
-dz\delta(d\bar{\theta})Y_{\bar{D}}D,
\end{align}
we find the transformation properties
\begin{align}
 \mu_{1|2}^0=&\mu_{1|2}^{0\,\prime}
-d\left[2\log\frac{D\theta'}{\bar{D}\theta'}
\,\mu_{0|2}^{0\,\prime}\right]
+\bm{\delta}_\mathrm{B}
\left[2\log\frac{D\theta'}{\bar{D}\bar{\theta}'}
\,\mu_{1|2}^{-1\,\prime}\right],
\\
 \mu_{2|2}^{-1}=&\mu_{2|2}^{-1\,\prime}
-d\left[2\log\frac{D\theta'}{\bar{D}\theta'}
\,\mu_{1|2}^{-1\,\prime}\right]
-\bm{\delta}_\mathrm{B}
\left[2\log\frac{D\theta'}{\bar{D}\bar{\theta}'}
\,\mu_{2|2}^{-2\,\prime}\right].
\end{align}

Moreover, noting that 
(\ref{eq:w_1|2^2}), (\ref{eq:rho_1|2^1}) and (\ref{eq:tilderho_1|2^1})
are rewritten by using (\ref{eq:star-def}) and (\ref{eq:tildestar-def}):
\begin{align}
\omega_{1|2}^2=&
dz\delta(d\theta)Y_{\bar{D}}
\left[\partial C\left\{
\frac{1}{2}(\bar{D}\tilde{C})-\tilde{C}\bar{D}\right\}-
\tilde{\mathcal C}^{(2)}
\right]
V
+d\bar{z}\delta(d\bar{\theta})
Y_D\left[
{\mathcal C}^{(2)}
-\bar{\partial}\tilde{C}\left\{
\frac{1}{2}(DC)-CD\right\}\right]V,
\\
 \rho_{1|2}^1=&
dz\delta(d\theta)Y_{\bar{D}}
\left[\partial C\left\{
\frac{1}{2}(\bar{D}\tilde{C})-\tilde{C}\bar{D}\right\}-
\tilde{\mathcal C}^{(2)}
\right]
YV
+d\bar{z}\delta(d\bar{\theta})Y_D
\left\{(\partial C-\bar{\partial}\tilde{C})
\frac{1}{2}C\delta(DC)V
\right\},
\\
 \tilde{\rho}_{1|2}^1=&
dz\delta(d\theta)Y_{\bar{D}}
\left\{
(\partial C-\bar{\partial}\tilde{C})\frac{1}{2}\tilde{C}
\delta(\bar{D}\tilde{C})V
\right\}
+d\bar{z}\delta(d\bar{\theta})
Y_D\left[
{\mathcal C}^{(2)}
-\bar{\partial}\tilde{C}\left\{
\frac{1}{2}(DC)-CD\right\}\right]\tilde{Y}V,
\end{align}
a straightforward but lengthy computation leads to
\begin{align}
 \omega_{1|2}^2=&
\omega_{1|2}^{2\,\prime}
-d\left[
2\log\frac{D\theta'}{\bar{D}\bar{\theta}'} \omega_{0|2}^{2\,\prime}
+\psi_{0|2}^{2\,(0,0)}\right]
+\bm{\delta}_\mathrm{B}\left[
2\log\frac{D\theta'}{\bar D \bar \theta'} \omega_{1|2}^{1\,\prime}
+\bar \psi_{1|2}^{1(0,0)}
-\psi_{1|2}^{1(0,0)}\right],
\\
 \rho_{1|2}^1=&
\rho_{1|2}^{1\,\prime}
-d\left[
2\log\frac{D\theta'}{\bar{D}\bar{\theta}'} \rho_{0|2}^{1\,\prime}
+\psi_{0|2}^{1\,(-1,0)}\right]
+\bm{\delta}_\mathrm{B}\left[
2\log\frac{D\theta'}{\bar D \bar \theta'} \rho_{1|2}^{0\,\prime}
-\psi_{1|2}^{1(0,0)}Y'\right],
\\
 \tilde{\rho}_{1|2}^1=&
\tilde{\rho}_{1|2}^{1\,\prime}
-d\left[
2\log\frac{D\theta'}{\bar{D}\bar{\theta}'} \tilde{\rho}_{1|2}^{0\,\prime}
+\psi_{0|2}^{1\,(0,-1)}\right]
+\bm{\delta}_\mathrm{B}\left[
2\log\frac{D\theta'}{\bar D \bar \theta'} \tilde{\rho}_{1|2}^{0\,\prime}
+\bar{\psi}_{1|2}^{1(0,0)}\tilde{Y}'\right],
\end{align}
where $\psi_{1|2}^{1(0,0)}$ and $\bar{\psi}_{1|2}^{1(0,0)}$ are
defined by
\begin{align}
 \psi_{1|2}^{1(0,0)}=dz'\delta(d\theta')Y_{\bar{D}}(\bar D \bar \theta')^{-2}
\bar \partial \bar \theta' 2\tilde C' V',
\quad
 \bar \psi_{1|2}^{1(0,0)}=d\bar z'\delta(d\bar \theta')
Y_D(D \theta')^{-2}
\partial \theta' 2C' V'.
\end{align}
By a similar computation,
we obtain the following transformations:
\begin{align}
 \omega_{2|2}^1=&
\omega_{2|2}^{1\,\prime}
-d\left[
2\log\frac{D\theta'}{\bar{D}\bar{\theta}'} \omega_{1|2}^{1\,\prime}
+\bar \psi_{1|2}^{1(0,0)}
-\psi_{1|2}^{1(0,0)}\right]
+\bm{\delta}_\mathrm{B}\left[
2\log\frac{D\theta'}{\bar D \bar \theta'} \omega_{2|2}^{0\,\prime}
\right],
\\
 \rho_{2|2}^0=&
\rho_{2|2}^{0\,\prime}
-d\left[
2\log\frac{D\theta'}{\bar{D}\bar{\theta}'} \rho_{1|2}^{0\,\prime}
-\psi_{1|2}^{1\,(-1,0)}Y'\right]
+\bm{\delta}_\mathrm{B}\left[
2\log\frac{D\theta'}{\bar D \bar \theta'} \rho_{2|2}^{-1\,\prime}
\right],
\\
 \tilde{\rho}_{2|2}^0=&
\tilde{\rho}_{2|2}^{0\,\prime}
-d\left[
2\log\frac{D\theta'}{\bar{D}\bar{\theta}'} \tilde{\rho}_{1|2}^{0\,\prime}
+\bar{\psi}_{1|2}^{1\,(0,0)}\tilde{Y}'\right]
+\bm{\delta}_\mathrm{B}\left[
2\log\frac{D\theta'}{\bar D \bar \theta'} \tilde{\rho}_{2|2}^{-1\,\prime}
\right].
\end{align}

In summary, the transformation laws of all elements in the descent sequence
are organized into a universal structure consisting of
$\bm{\delta}_{\mathrm B}$-exact and $d$-exact terms.
This guarantees that the entire descent sequence
is well-defined in BRST cohomology
and transforms consistently under superconformal transformations.

\section{Concluding remarks
\label{sec:remarks}}

In this paper, we have developed a geometric framework
for the construction of vertex operators in superstring theory
based on integral forms on super Riemann surfaces.
This framework provides a systematic and unified description
of vertex operators across different ghost and picture sectors.
Starting from the top integral form representing the integrated
NS-NS vertex operator, we derived a class of descent equations
and constructed vertex operators with various ghost and picture numbers.
By incorporating inverse picture-changing operators,
we extended this construction to a class of picture sectors,
and further generalized it to higher ghost number operators
by introducing the superfield analogue of
$\partial c-\bar{\partial}\tilde c$.
We have shown that, although additional contributions are required
in the superstring case, all resulting operators are organized into
a consistent descent structure and are well-defined in BRST cohomology.
In particular, they satisfy a universal descent equation of the form
\begin{align}
\bm{\delta}_\mathrm{B}\omega_{m|n}^g
=
d\omega_{m-1|n}^{g+1}.
\end{align}
The same descent structure extends to the sequences obtained by acting
with inverse picture-changing operators, such as $\rho$, $\tilde{\rho}$,
and $\mu$, demonstrating the universality of the construction across
different picture sectors.

An important outcome of our analysis is a precise correspondence between
geometric objects on the parity-reversed tangent bundle and ghost
superfields. In contrast to the bosonic case, where the relation
$dz \leftrightarrow c$ provides the geometric origin of ghost insertions,
the superstring case requires a refined identification at the level of
superfields. In particular, the natural one-form $dz-\theta d\theta$ and
the even differential $d\theta$ are related to the ghost superfield and
its superderivative through the correspondence (\ref{eq:dzdthetaC}),
which provides a geometric realization of the superghost structure.
This correspondence underlies the structure of inverse picture-changing
operators and clarifies the geometric origin of superghost insertions
and their role in the descent construction.

One important issue concerns the scope of picture numbers captured
in the present framework. Although we have incorporated the inverse
picture-changing operator $Y$ to construct vertex operators in certain
picture sectors, our analysis does not provide a description of
operators at arbitrary picture numbers within the descent equation
framework. To achieve a complete characterization, it is necessary to
reconsider the picture-changing operation of \cite{Friedan:1985ge}
within the superfield formalism and to formulate it consistently in
terms of descent equations. This remains an important problem for future
investigation.

A further extension is the incorporation of the Ramond sector. As
emphasized by Witten \cite{Witten:2012ga},
Ramond vertex operators are associated with Ramond
punctures in supergeometry, which are qualitatively different from NS
punctures. In particular, Ramond vertices cannot be described simply in
terms of superfields on the super Riemann surface. It is therefore an
important open problem to incorporate Ramond punctures into the
present formulation and to extend the geometric framework accordingly.

Another important problem is 
the extension of the present framework to 
the zero-momentum dilaton in superstring theory.
In our
previous analysis \cite{Kishimoto:2024yuw} in the bosonic string, it was
shown that, for the zero-momentum dilaton, the naive vertex operator
must be modified by additional terms involving the ghost fields and the
target-space coordinates $X^\mu$ in order to cancel quantum corrections
and to satisfy the descent equations consistently. At ghost number two,
the unintegrated vertex operator is corrected by the well-known ghost
dilaton term, while at ghost number three, a novel contribution appears,
as clarified in our previous work.
This additional term plays an
essential role: the disk one-point amplitude of the dilaton can be
correctly reproduced within the BRST formalism only when this term is
included.
It is therefore an important open problem to understand
how such additional contributions
arise and are organized within the framework of integral forms and
descent equations.

\section*{Acknowledgments}
We would like to thank H.~Kunitomo for valuable discussion and comments.
S.~S.~was supported in part
by JSPS Grant-in-Aid for Scientific
Research (C) \#22K03625. H.~S.~was supported in part by
JST SPRING, Grant Number JPMJSP2115. T.~T.~was supported in part by
JSPS Grant-in-Aid for Scientific Research (C) \#23K03388.

\bibliographystyle{utphys}
\bibliography{referencev4}

@Article{Hata1987b,
  author       = {Hata, Hiroyuki and Itoh, Katsumi and Kugo, Taichiro and Kunitomo, Hiroshi and Ogawa, Kaku},
  title        = {{BRS INVARIANT VERTEX OF NEVEU-SCHWARZ-RAMOND SUPERSTRING}},
  journal      = {Prog. Theor. Phys.},
  year         = {1987},
  volume       = {78},
  pages        = {453},
  doi          = {10.1143/PTP.78.453},
  slaccitation = {%%CITATION = PTPKA,78,453;%%},
}

@Article{Thorn1989,
  author    = {Thorn, Charles B.},
  title     = {{STRING FIELD THEORY}},
  journal   = {Phys.Rept.},
  year      = {1989},
  volume    = {175},
  pages     = {1-101},
  doi       = {10.1016/0370-1573(89)90015-X},
}

@Article{Witten1986a,
  author    = {Witten, Edward},
  title     = {{Interacting Field Theory of Open Superstrings}},
  journal   = {Nucl.Phys.},
  year      = {1986},
  volume    = {B276},
  pages     = {291},
  doi       = {10.1016/0550-3213(86)90298-1},
}

@article{Friedan:1985ge,
      author         = "Friedan, Daniel and Martinec, Emil J. and Shenker,
                        Stephen H.",
      title          = "{Conformal Invariance, Supersymmetry and String Theory}",
      journal        = "Nucl. Phys.",
      volume         = "B271",
      year           = "1986",
      pages          = "93-165",
      doi            = "10.1016/0550-3213(86)90356-1,
                        10.1016/S0550-3213(86)80006-2",
      reportNumber   = "PRINT-86-0024 (CHICAGO), EFI-85-89-CHICAGO",
      SLACcitation   = "%%CITATION = NUPHA,B271,93;%%"
}

@book{Polchinski:1998rr,
      author         = "Polchinski, J.",
      title          = "{String theory. Vol. 2: Superstring theory and beyond}",
      publisher      = "Cambridge University Press",
      year           = "2007",
      ISBN           = "9780511252280, 9780521633048, 9780521672283",
      SLACcitation   = "%%CITATION = INSPIRE-487241;%%"
}

@article{Belopolsky:1997bg,
    author = "Belopolsky, Alexander",
    title = "{New geometrical approach to superstrings}",
    eprint = "hep-th/9703183",
    archivePrefix = "arXiv",
    reportNumber = "IFP-103-UNC",
    month = "3",
    year = "1997"
}

@article{Belopolsky:1996cy,
    author = "Belopolsky, Alexander",
    title = "{De Rham cohomology of the supermanifolds and superstring BRST cohomology}",
    eprint = "hep-th/9609220",
    archivePrefix = "arXiv",
    reportNumber = "IFP-101-UNC",
    doi = "10.1016/S0370-2693(97)00445-0",
    journal = "Phys. Lett. B",
    volume = "403",
    pages = "47--50",
    year = "1997"
}

@article{Belopolsky:1997jz,
    author = "Belopolsky, Alexander",
    title = "{Picture changing operators in supergeometry and superstring theory}",
    eprint = "hep-th/9706033",
    archivePrefix = "arXiv",
    reportNumber = "IFP-104-UNC",
    month = "6",
    year = "1997"
}

@article{Witten:2012ga,
    author = "Witten, Edward",
    title = "{Notes On Super Riemann Surfaces And Their Moduli}",
    eprint = "1209.2459",
    archivePrefix = "arXiv",
    primaryClass = "hep-th",
    doi = "10.4310/PAMQ.2019.v15.n1.a2",
    journal = "Pure Appl. Math. Quart.",
    volume = "15",
    number = "1",
    pages = "57--211",
    year = "2019"
}

@article{Witten:2012bg,
    author = "Witten, Edward",
    title = "{Notes On Supermanifolds and Integration}",
    eprint = "1209.2199",
    archivePrefix = "arXiv",
    primaryClass = "hep-th",
    doi = "10.4310/PAMQ.2019.v15.n1.a1",
    journal = "Pure Appl. Math. Quart.",
    volume = "15",
    number = "1",
    pages = "3--56",
    year = "2019"
}

@article{Kishimoto:2024yuw,
    author = "Kishimoto, Isao and Kouga, Mako and Seki, Shigenori and Takahashi, Tomohiko",
    title = "{Closed string vertex operators with various ghost number}",
    eprint = "2402.06179",
    archivePrefix = "arXiv",
    primaryClass = "hep-th",
    doi = "10.1016/j.nuclphysb.2024.116549",
    journal = "Nucl. Phys. B",
    volume = "1004",
    pages = "116549",
    year = "2024"
}

@article{Yamron:1986nb,
    author = "Yamron, Jonathan P.",
    title = "{A Gauge Invariant Action for the Free Ramond String}",
    reportNumber = "UCB-PTH-86/3",
    doi = "10.1016/0370-2693(86)91131-7",
    journal = "Phys. Lett. B",
    volume = "174",
    pages = "69--74",
    year = "1986"
}

@article{Voronov:1986a,
    author = "Voronov, T and Zorich, A",
    title = "{Integral transformations of pseudodifferential forms}",
    reportNumber = "",
    doi = "",
    journal = "Usp. Mat. Nauk.",
    volume = "41",
    pages = "167",
    year = "1986"
}

@article{Voronov:1986b,
    author = "Voronov, T and Zorich, A",
    title = "{Complex of forms on a supermanifold}",
    reportNumber = "",
    doi = "",
    journal = "Funkc. Anal. Prilozh.",
    volume = "20",
    pages = "58",
    year = "1986"
}

@article{Voronov:1987,
    author = "Voronov, T and Zorich, A",
    title = "{Theory of bordisms and homotopy properties of supermanifolds}",
    reportNumber = "",
    doi = "",
    journal = "Funkc. Anal. Prilozh.",
    volume = "21",
    pages = "77",
    year = "1987"
}

@article{Voronov:1988,
    author = "Voronov, T and Zorich, A",
    title = "{Cohomology of supermanifolds and integral geometry}",
    reportNumber = "",
    doi = "",
    journal = "Sov. Math. Dokl.",
    volume = "37",
    pages = "96",
    year = "1988"
}

\end{document}